\documentclass[preprint,3p,times,twocolumn]{elsarticle}

\usepackage{hyperref}
\usepackage{lineno}
\usepackage{amsmath}
\usepackage{xcolor}
\usepackage{upgreek}
\usepackage{booktabs}

\modulolinenumbers[5]

\makeatletter
\def\ps@pprintTitle{%
  \let\@oddhead\@empty
  \let\@evenhead\@empty
  \def\@oddfoot{\reset@font\hfil\thepage\hfil}
  \let\@evenfoot\@oddfoot
}
\makeatother

\begin{document}

\begin{frontmatter}

\title{Energy Optimal Point-to-Point Motion Profile Optimization}

\author[1,4]{Nick Van Oosterwyck \texorpdfstring{\corref{cor1}}} \ead{nick.vanoosterwyck@uantwerpen.be}
\author[2,5]{Foeke Vanbecelaere \texorpdfstring{\fnref{fn1}}} \ead{foeke.vanbecelaere@ugent.be}
\author[3]{Ferre Knaepkens} \ead{ferre.knaepkens@uantwerpen.be}
\author[2,5]{Michael Monte} \ead{michael.monte@ugent.be}
\author[2,5]{Kurt Stockman} \ead{kurt.stockman@ugent.be}
\author[3,6]{Annie Cuyt} \ead{annie.cuyt@uantwerpen.be}
\author[1,4]{Stijn Derammelaere} \ead{stijn.derammelaere@uantwerpen.be}

\cortext[cor1]{Corresponding author}
\fntext[fn1]{Equal contribution}
\address[1]{Department of Electromechanics, CoSys-Lab, University of Antwerp, Groenenborgerlaan 171, 2020 Antwerp, Belgium}
\address[2]{Department of Electrical Energy, Metals, Mechanical Constructions and Systems, Ghent University campus Kortrijk, 8500 Kortrijk, Belgium}
\address[3]{Department of Mathematics and Computer Science, University of Antwerp, Middelheimlaan 1, 2020 Antwerpen, Belgium}
\address[4]{AnSyMo/CoSys, Flanders Make, the strategic research center for the manufacturing industry}
\address[5]{FlandersMake@UGent - Core Lab EEDT-MP}
\address[6]{College of Mathematics and Statistics, Shenzhen University Shenzhen, Guangdong 518060, China}




\begin{abstract}
Position-controlled systems driving repetitive tasks are of significant importance in industrial machinery. The electric actuators used in these systems are responsible for a large part of the global energy consumption, indicating that major savings can be made in this field. In this context, motion profile optimization is a very cost-effective solution as it allows for more energy-efficient machines without additional hardware investments or adaptions. In particular, mono-actuated mechanisms with position-dependent system properties have received considerable attention in literature. However, the current state-of-the-art methods often use unbounded design parameters to describe the motion profile. This both increases the computational complexity and hampers the search for a global optimum. In this paper, Chebyshev polynomials are used to describe the motion profile. Moreover, the exact bounds on the Chebyshev design parameters are derived. This both seriously reduces the computational complexity and limits the design space, allowing the application of a global optimizer such as the genetic algorithm. Experiments validate the added value of the chosen approach. In this study, it is found that the energy consumption can be reduced by 62.9\% compared to a standard reference motion profile.
\end{abstract}

\begin{keyword}
Motion profile optimization \sep Point-to-point motion \sep Energy efficiency \sep Validation \sep CAD model
\MSC[2020] 49
\end{keyword}

\end{frontmatter}


\section{Introduction}
In the last decades, economic considerations and stricter government regulations have driven engineers to come up with new techniques to reduce the energy consumption of industrial machinery. Statistics indicate that electric motors are generally responsible for about 2/3 of the industrial electricity consumption, which indicates that major savings are to be made in this field \cite{Bo2008}.

\begin{figure}[thpb] 
  \centering	
  \includegraphics[width=\columnwidth]{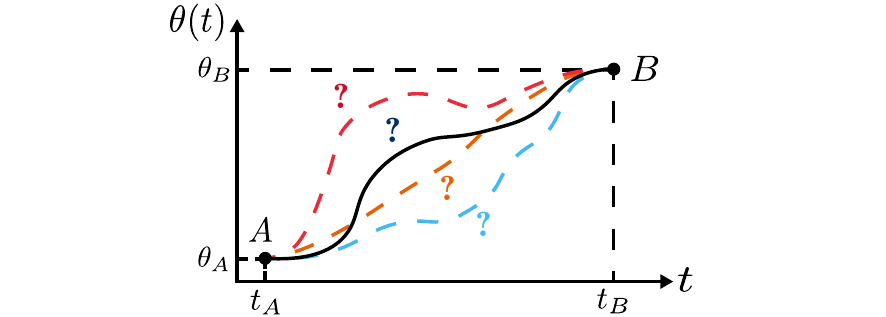}
  \caption{Motion profile of a PTP movement with constraints $\theta_A$, $\theta_B$, $t_A$ and $t_B$.}
  \label{fig:generalprofile}
\end{figure}

In this context, several technologies and methods have been developed to reduce the electrical energy consumption of mechatronic systems. So-called, motion profile optimization starts from the idea that in many industrial applications, only part of the motion is constrained by the process requirements. Hence, an optimization potential rises in the non-constrained part of the position function $\theta(t)$, in between the start and endpoint of the point-to-point (PTP) motions (Fig \ref{fig:generalprofile}).

By modifying the position function $\theta(t)$ between a fixed start- $\theta(t_A) = \theta_A$ and endpoint $\theta(t_B) = \theta_B$, it is possible to minimize one or several design objectives such as energy \cite{VanOosterwyck2020}, motion time  \cite{Richiedei2016,Ceulemans2021}, machine cost \cite{Derammelaere2017b} and vibrations \cite{Lee2020}. Moreover, since these industrial applications often involve repetitive movements, the motion profile optimization effect will be perceptible every machine cycle, thus, making it an indispensable step in modern energy-efficient machine design. 

    
    
    
    

\subsection{Related work}


Dedicated machines with single-axis servo mechanisms are omnipresent in industrial production processes, especially given the tendency to evolve towards dedicated actuators for each machine movement \cite{Berselli2016}. In this context, several approaches have been presented in the past literature to optimize the motion profiles of these one DOF systems.

In \cite{Richiedei2016,Lee2020,Park1996,Carabin2021,Botan2010}, optimal motion profiles are obtained for mechanical systems with constant load parameters such as the inertia $J$. Nevertheless, as indicated in literature \cite{Pellicciari2015,Berselli2016,Hsu2014}, it is essential to consider varying loads to cover the majority of machine applications.

However, in those cases obtaining an analytic description of the position-dependent system properties is a challenging task for industrial machine designers. For instance, \cite{Hsu2014,Hsu2016,Huang2012,Ha2006} use Hamilton's principle and Lagrange multipliers to obtain the differential-algebraic equations of the system. In \cite{Sollmann2010}, a method of virtual work is described to obtain the system matrix while \cite{Vanbecelaere2020} determines the inertia profile using the method of kinetic energy. Such approaches are cumbersome, complex and error-prone, and are not easily applicable. Especially given the trend indicated in \cite{Walsch2014} that there is a demand for methods that take into account the ease of implementation.

Fortunately, machine builders often already design their machines in 3D CAD multibody software, which can be used to extract crucial information. Hence, in \cite{VanOosterwyck2019} and \cite{Berselli2016}, the authors of this paper describe a technique to derive the position dependency of critical parameters inertia $J(\theta)$ and load torque $\tau_l(\theta)$, based on only three CAD motion simulations. If the load model is known, optimization algorithms can iterate on it to minimize energy usage.

For what concerns these optimization algorithms, several approaches are possible. On the one hand, \cite{Park1996} and \cite{Shiller1996} use an indirect approach such as Pontryagin's Maximum Principle to obtain the best possible control. However, this method tends to be abandoned recently due to the small convergence area and difficulties incorporating constraints \cite{Chettibi2004}.

On the other hand, direct approaches recast the optimization into a nonlinear optimization problem, which can be solved with different numerical methods. In particular, \cite{VanOosterwyck2020} and \cite{Pellicciari2015} use gradient-based methods such as Sequential Quadratic Programming (SQP) or quasi-Newton methods that are known to have very low solve times and good scalability. However, these algorithms can only deliver local optimal solutions and are not suited for problems with multiple minima. As indicated in \cite{Huang2017}, the optimum obtained with gradient-based methods is greatly influenced by the selected starting points, which are to be chosen arbitrarily.

To avoid this problem, heuristic optimization algorithms \cite{VanOosterwyck2019,Huang2012,Hsu2014,Berselli2016}  such as generalized pattern search (GPS) or genetic algorithms (GA) are of interest. In contrast to gradient-based algorithms that do not search the entire design space, derivative-free algorithms like GA often sample a wide part of the design space in order to be successful \cite{Wenzhong2005}. Nevertheless, because these heuristic solvers do not exploit gradient information, they are not computationally competitive with gradient-based methods \cite{Betts1998}.

Regarding the motion profile function, several papers rely on piecewise position functions (\cite{Pellicciari2015,Berselli2016}), where either cubic, quintic, or trigonometric splines are used. However, the objective functions in these works, are characterized by many local minima, causing the risk of getting stuck in a suboptimal solution. For instance, in \cite{Piazzi1998}, the usage of cubic splines resulted in a savings difference of $18\%$ between the global and local optimum.

On the other hand, continuous motion profile functions such as classic polynomials (\cite{VanOosterwyck2019,Huang2012,Hsu2014,Carabin2021,Lee2020}) are also popular because they do not introduce high jerk peaks into the system, which increases the wear of the components. However, the resulting optimization problem is known to be badly conditioned. For example in \cite{VanOosterwyck2019}, the coefficients reached values up to $1.8 \, 10^{20}$. Therefore, the authors of this paper proposed Chebyshev polynomials and rescalings in \cite{VanOosterwyck2020} to obtain a more numerically stable problem formulation and to increase the robustness against getting stuck in local minima. Moreover, in contrast to the classical polynomial descriptions, the design variables of Chebyshev polynomials can be bounded as is shown in this paper.

\subsection{Method}
In light of the considerations mentioned above, a CAD-based method for computing energy-optimal PTP (Point-to-Point) motion profiles of single DOF mechanisms using Chebyshev polynomials has been previously presented by the authors in \cite{VanOosterwyck2020}. However, due to the numerous symbolic calculations involved in constructing the objective function, solve times of almost 2 hours were reported. In addition, the solutions in \cite{VanOosterwyck2020} were obtained using gradient-based solvers which have a high risk of getting stuck in local minima. Finally, only theoretical reductions were reported, thus, leaving the feasibility of the proposed motion profiles undetermined. This paper builds upon these previous results by providing five critical improvements:

\begin{itemize}

    \item In order to reduce the computational burden, a discrete approach is presented which eliminates the use of symbolic operations. To do so, the discrete system property data which originates from the CAD motion simulations have to be properly rescaled and interpolated.
    
    \item As an accurate model of the system dynamics is crucial for a correct optimization, the dynamics of the mechanism are extended by including damping and friction into the optimization routine. In addition, a new identification procedure is described which is validated on an industrial case.
    
    \item A derivation for exact bounds of the Chebyshev polynomial coefficients is introduced. This allows limiting the feasible design space. The latter is essential for heuristic optimizers to reduce their computation time and the chances that the global optimum remains unidentified.
    
    \item To check the robustness of the proposed method against getting stuck in local optima, the resulting optimization problem is solved with both a fast gradient-based and a global heuristic solver (i.e. GA). 
    
    \item Experimental tests have been carried out on an industrial pick-and-place unit to quantify the actual measured energy savings.
    
   
\end{itemize}



\section{System Modeling}
\begin{figure}[thpb]
  \centering	
  \includegraphics[width=\columnwidth]{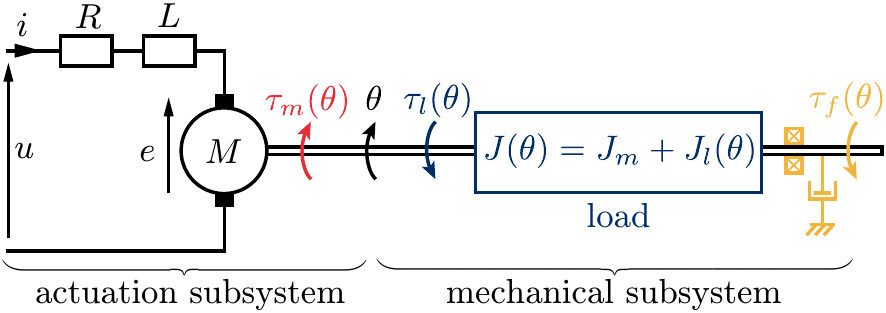}
  \caption{Schematic of the q-axis of a single axis mechanism.}
  \label{fig:Simplified_Model}
\end{figure}

The complete mechatronic system can be divided into two subsystems (Fig. \ref{fig:Simplified_Model}). On the one hand, there is the \textit{mechanical subsystem} which describes the dynamics of a generic single-axis system. For high dynamical applications, these systems usually consist of slider-crank mechanisms and four-bar linkages \cite{Berselli2016}. Nevertheless, the approach is valid to any position-controlled system where the mechanism is driven by a single actuator.

On the other hand, there is the \textit{actuation subsystem} which converts the electrical energy into mechanical energy and drives the mechanism. For the envisaged position-controlled systems, PMSMs are becoming the industry standard for rotary applications, whereas linear motors are used for fast and precise linear movements \cite{Kiel2008}. In Fig. \ref{fig:Simplified_Model}, the PMSM actuator is represented by an equivalent DC model.


\subsection{Mechanical subsystem}
The dynamics of a single axis DOF mechanism can be described by means of the torque equation \cite{Pellicciari2015,Dresig2010} :

\begin{equation} \label{eq:torque_equation}
\tau_m(t) = \tau_l(\theta) + \underbrace{J(\theta)\ddot{\theta}}_{\tau_a} + \underbrace{\frac{1}{2}\frac{\mathrm{d}J(\theta)}{\mathrm{d}\theta}(\dot{\theta})^2}_{\tau_{v}} +  \tau_f(\dot{\theta}) \, .
\end{equation}

With reference to Fig. \ref{fig:Simplified_Model} and equation (\ref{eq:torque_equation}), let us define $\theta = \theta(t)$ as the Lagrangian coordinate which describes the angular position of the main driving axis as a function of time $t$. The motor torque $\tau_m(\theta)$ is defined as the driving torque generated by the motor. The load torque $\tau_l(\theta)$ contains both gravitational forces as well as external process powers that act on the mechanism.

Furthermore, all inertias of the mechanism's components are related to the main driving axis resorting to the concept of reduced moment of inertia. Therefore, the reduced inertia of the complete system $J(\theta)$ is defined as a combination of the reduced load inertia $J_l(\theta)$ and inertia of the motor shaft itself $J_m$. Note that the position-dependent inertia of the system $J(\theta)$, results in two torque components when it is reduced to the motor side. The acceleration torque $\tau_a$ represents the part of the motor torque responsible for the motor acceleration forces that arise during the movement, while the variation torque $\tau_v$ compensates for the variation of inertia in the system. 

Finally, the frictional torque $\tau_f(\dot{\theta})$ is defined as the result of frictional forces such as, for instance, viscous brush friction or dry bearing friction in the motor bearings and mechanical system. A commonly used model of friction shows three components of force: Coulomb (sliding) friction, viscous damping, and static friction \cite{Ellis2012}. Regarding the PMSM, as indicated in \cite[p.~175]{Westphal2001}, the only appreciable friction effect in operation is viscous friction. Thus, coulomb and static frictions can be neglected in the PMSM model. For what concerns the mechanical model, only the viscous damping is modeled since the other friction components are constant and will not have an effect on the optimal motion profile:

\begin{equation} \label{eq:driction_torque}
\tau_f(\dot{\theta}) = \mu_v\dot{\theta} \, ,
\end{equation}

with $\mu_v$ the equivalent viscous friction coefficient. 

The key benefit of the formulation in \eqref{eq:torque_equation} is that it permits to model every possible mechanism with a known geometry and allows to define a generic optimization approach.

\subsection{Actuation subsystem}
Concerning the dynamics of the PMSM as depicted in Fig. \ref{fig:Simplified_Model} (represented by an equivalent DC model), the electromechanical behavior can be easily described by the following basic laws \cite[p.~843]{Rizzoni2003}:

\begin{equation} \label{eq:basis_torque}
\tau_m = k_t i \, ,
\end{equation}



\begin{equation} \label{eq:voltage}
u = R i + L\frac{\mathrm{d}i}{\mathrm{d}t} + e = R i + L\frac{\mathrm{d}i}{\mathrm{d}t} + p k_v \dot{\theta} \, ,
\end{equation}

with electric back emf $e$, resistance $R$, back emf constant $k_v$, motor torque constant $k_t$, and number of pole pairs $p$, which can be found in the motor data sheet. 

In equation (\ref{eq:voltage}), the voltage drop $L\frac{\mathrm{d}i}{\mathrm{d}t}$ due to the armature inductance is omitted as the mean value of its reactive power will be zero and therefore does not contribute to the system's energy need \cite{Pellicciari2015}.



Depending on whether the electric power flows from the drive unit to the PMSM's or vice versa, the PMSM operates in respectively motor or generator mode. In this latter condition, depending on the capabilities of the drive unit, the generated electric power can be either stored in a capacitor, dissipated as heat on a braking resistance, or transferred back to the energy source. Recent commercial PMSM drives are sized so that no electric power is actually dissipated during normal functioning so that the braking resistance is actually activated only under emergency conditions \cite{Berselli2016}. Therefore, in what follows, it is assumed that all the generated energy is returned to the grid and no losses occur in the process.

For a correct model of the actuation subsystem and prediction of the energy usage, it is important to model other losses such as cooling fans and drive circuitry as well. Nevertheless, the power consumption of these devices is generally considered constant and is therefore not affected by the motion profile \cite{Gadaleta2019}.

In order to minimize the total energy need $E$ of the application, it is crucial to quantify the input energy of the complete system. Therefore, similar to \cite{Berselli2016}, a formulation of the input electrical energy $E$ is derived and a torque-based design objective is obtained which allows minimizing the energy solely based on the mechanical parameters.

Starting from equations \eqref{eq:basis_torque} and \eqref{eq:voltage}, the instantaneous power $P_e$ is defined as

\begin{equation}
P_{e} = u \, i = \frac{R}{k_t^2}\tau_m^{2} + \frac{p k_v}{k_t} \tau_m \dot{\theta} \, .
\end{equation}

The motion profile is defined on the time interval $t \in [t_A,t_B]$ and must have zero initial and final speed and acceleration, i.e. $\dot{\theta}(t_A) = \dot{\theta}(t_B) = \ddot{\theta}(t_A) = \ddot{\theta}(t_B) = 0$. The total energy can be expressed as

\begin{equation}
\begin{aligned}
E= \int_{t_A}^{t_B} P_e \, \mathrm{d} t &= \int_{t_A}^{t_B} \left[ \frac{R}{k_t^2}\tau_m^{2} + \frac{p k_v}{k_t} \tau_m \dot{\theta} \right] \, \mathrm{d} t \, .
\end{aligned}
\end{equation}

Then, by incorporating the torque equation from \eqref{eq:torque_equation}, the total energy of the motion is given by:

\begin{equation} \label{eq:energy}
\begin{aligned}
E=\frac{p k_v}{k_t}  \underbrace{\int_{t_A}^{t_B} (\tau_a + \tau_v) \dot{\theta} \, \mathrm{d} t}_{E_{k}} + \frac{p k_v}{k_t} \underbrace{\int_{t_A}^{t_B} \tau_l \, \dot{\theta} \, \mathrm{d} t}_{E_{p}} \\
+\underbrace{\int_{t_A}^{t_B}\left[\frac{R}{k_t^2} \tau_m^{2}+ \frac{p k_v}{k_t} \tau_f \dot{\theta}\right] \mathrm{d} t}_{E_{l}} \, .
\end{aligned}
\end{equation}

Here, the first term $E_k$ represents the kinetic energy of the moving masses  in the system. Due to the rest-to-rest motion of the envisaged applications, this term reduces to zero. Further, the term $E_p$ represents the potential energy stored in the system. As this term $E_p$ only depends on the fixed start $\theta_A$ and end position $\theta_B$, it is disregarded in the optimization routine \cite{Berselli2016}. The final term $E_l$ represents the energy that is lost due to the coil resistance and frictional forces and is the only term that is affected by optimizing the motion profile $\theta(t)$. Nevertheless, in many industrial applications, the frictional forces $\tau_f$ are negligible \cite{Park1996}, especially if the inertial loads are predominant. Thus, the energy losses $E_l$ can be expressed as:

\begin{equation}
    E_l = \int_{t_A}^{t_B} \frac{R}{k_t^2} \tau_m^{2} \mathrm{d} t = \frac{R \Delta t}{k_t^2}\tau_{rms}^2 \, ,
\end{equation}

where $\tau_{rms}$ is the RMS value of the motor Torque $\tau_{m}$. This proves that the RMS torque $\tau_{rms}$ can be effectively used as an optimization objective to minimize the total energy usage of the system. This is very useful in situations where the motor coil properties are unknown or where parameters are missing \cite{Berselli2016}.

\section{Identification}
\subsection{Inertia and Load Torque}
\begin{figure}[thpb]
  \centering	
  \includegraphics[width=\columnwidth]{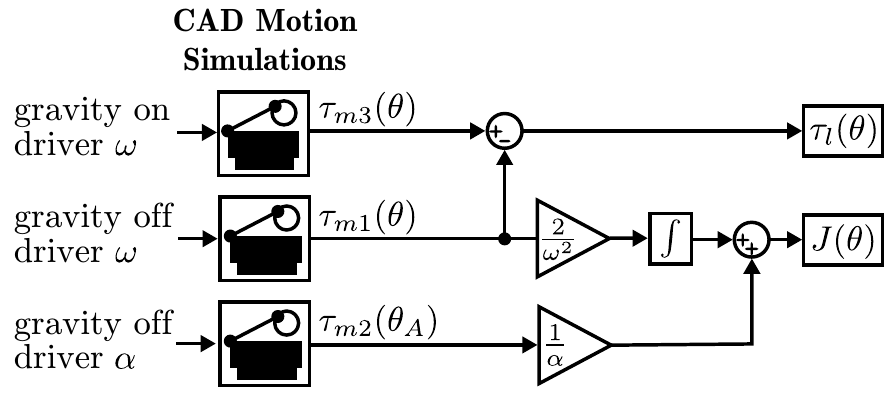}
  \caption{Schematic overview of the procedure for extracting position-dependent properties $J(\theta)$ and $\tau_l(\theta)$ based on three different CAD motion simulations \cite{VanOosterwyck2019}.}
  \label{fig:Property_Extraction}
\end{figure}

Identification of all the position varying parameters in the highly nonlinear differential torque equation (\ref{eq:torque_equation}) is not straightforward. Fortunately, machine builders design their machines in 3D CAD multibody software. For this reason, \cite{Berselli2016} and \cite{VanOosterwyck2019} describe a technique to derive the position dependency of critical parameters inertia $J(\theta)$ and load torque $\tau_l(\theta)$, based on three CAD motion simulations. (Fig. \ref{fig:Property_Extraction}). 

\begin{figure}
\centering
\includegraphics[width=0.47\columnwidth]{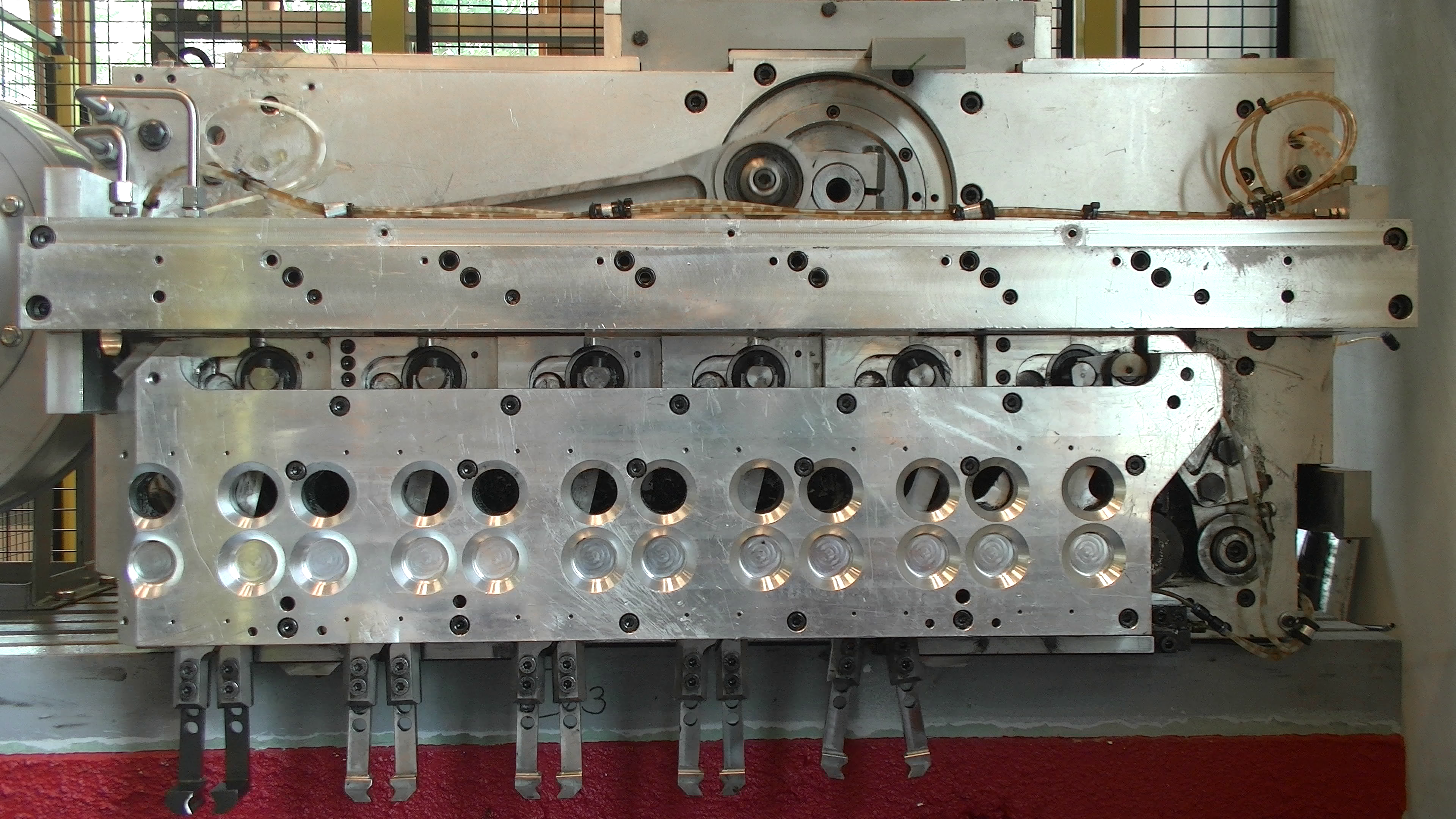}
\includegraphics[width=0.51\columnwidth]{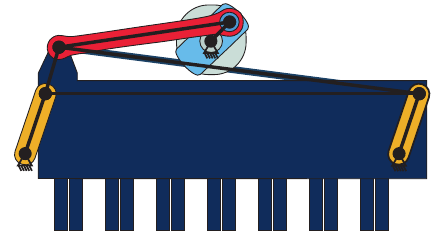}
\caption{Experimental set-up (left) and schematic overview (right) of the pick- and place unit.}
\label{fig:Experimental_Setup}
\end{figure}

\begin{figure}[thpb]
  \centering	
  \includegraphics[width=\columnwidth]{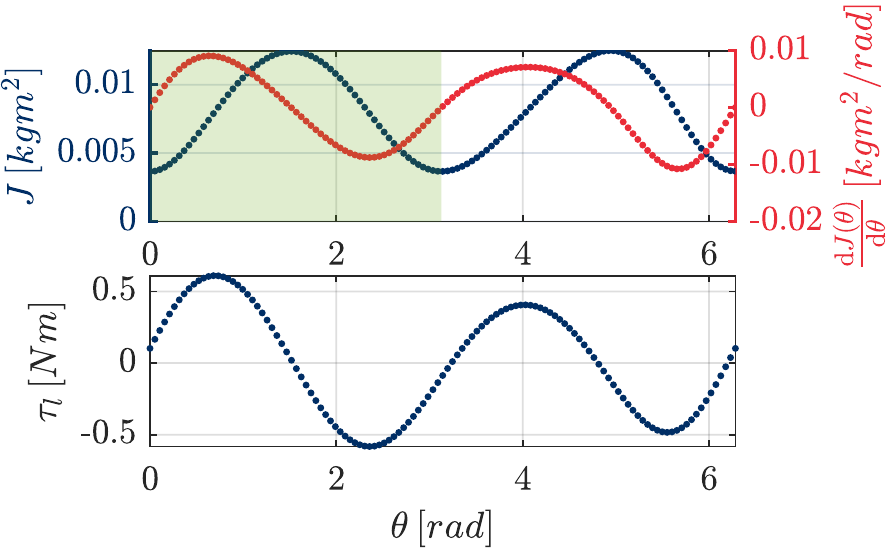}
  \caption{Values of system properties inertia $J(\theta)$ and load torque $\tau_l(\theta)$.}
  \label{fig:sytem_properties}
\end{figure}

In this paper, the identification routine is illustrated by applying it to an industrial pick-and-place unit (Fig. \ref{fig:Experimental_Setup}) that performs repetitive movements between start point $A$ with angular position $\theta_A = 0$ and endpoint $B$ with angular position $\theta_B = 173.6^\circ$. The resulting inertia $J(\theta)$ and load torque $\tau_l(\theta)$ profiles are presented in Fig. \ref{fig:sytem_properties}. Because of the machine position limits $\theta_A$ and $\theta_B$, only the green shaded part of the system properties is relevant during operation.

\subsection{Viscous Friction Coefficient}
Once the system properties $J(\theta)$ and $\tau_l(\theta)$ are determined, the only indefinite term in Eq. \eqref{eq:torque_equation} is the friction torque $\tau_f$, and more specifically $\mu_k$. In the previous description of the energy flows, the friction torque $\tau_f$ was neglected, leading to a simple objective (i.e. $\tau_{rms}$) to quantify the energy consumption. However, it is important to verify this statement for the intended setup. Therefore this section describes a method to quantify the frictional forces $\tau_f$.

Since the viscous friction coefficient $\mu_k$ parameter is highly dependent on the practical setup, it is often only possible to determine this parameter experimentally. Therefore, a first measurement is carried out by using an arbitrary motion profile $\theta^{*}(t)$ as a set point and recording the resulting actual motor torque $\tau_{m}^{e}(t)$ and position $\theta^{e}(t)$. The arbitrary motion profile $\theta^{*}(t)$ can be determined by using a default motion law such as a trapezoidal or s-curve profile. 

After this measurement, a least squares fit can be used to determine the experimental value of $\mu_k$, by fitting the torque model $\tau_{m}(\theta^{e}(t),\mu_k)$ with the measured torque $\tau_{m}^{e}$.

However, using the measured position $\theta^{e}$ and its time derivatives $\dot{\theta}^{e}$, $\ddot{\theta}^{e}$ in the torque equation \eqref{eq:torque_equation} leads to unfeasible results since the derivatives amplify any noise that is present in the measurement. Therefore, the measured position $\theta^{e}(t)$ is fitted with an $n$-th degree polynomial $\theta^p(t) = \sum_{i=1}^{n} a_it^i$ and differentiated symbolically to smooth out any noise. 


The friction parameter $\mu_v$ is thus determined by comparing the measured torque $\tau_{m}^{e}(t)$ with virtual model and fitted motion profile $\tau_{m}(\theta^{p}(t))$:

\begin{equation}
\begin{aligned}
& \underset{\mu_v \, \in \, \mathbb{R}}{\text{minimize}}
& & ||\tau_{m}^{e}(t) - \tau_{m}(\theta^{p}(t),\mu_v)||_2 \, .
\end{aligned}
\end{equation}

\begin{figure}[thpb]
  \centering	
  \includegraphics[width=\columnwidth]{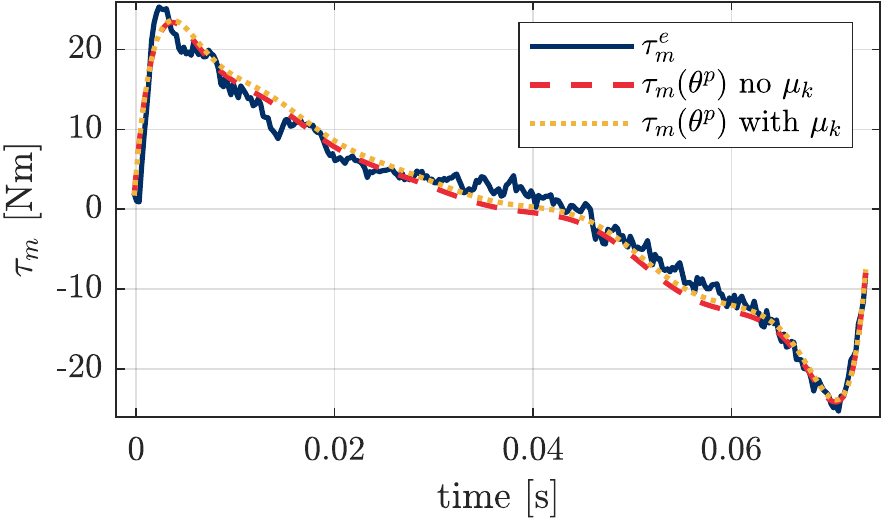}
  \caption{Comparison of the virtual $\tau_{m}(\theta^{p}(t))$ and measured torque $\tau_{m}^{e}(t)$ (with and without friction).}
  \label{fig:Torque_Validation}
\end{figure}

For the pick-and-place unit, a viscous damping coefficient of $0.0157Nms/rad$ was found. In Fig. \ref{fig:Torque_Validation}, a comparison of the measured $\tau_{m}^{e}(t)$ and virtual $\tau_{m}(\theta^{p}(t))$ torque is presented. The difference between the virtual torques with and without friction is minimal, which indicates that the friction can be neglected for the present case. The graph also shows a close correlation between the virtual and measured torque, which indicates that the virtual model can be effectively used to minimize the RMS torque $\tau_{rms}$ and, by extension, the energy consumption $E$. 

\section{Optimization Approach}
\subsection{Motion Profile Definition \& Rescaling}
In this paper, a Chebyshev polynomial $\sum_{i=0}^{n} p_iT_i(x)$ is used to define the position profile $\theta(t)$, where $t \in [t_A,t_B]$, in between the start- ($\theta(t_A) = \theta_A$) and endpoint ($\theta(t_B) = \theta_B$) of the motion task. The sequence of orthogonal Chebyshev polynomials $T_k(x) = T_k(\cos(\vartheta))$, defined on the interval $x \in [-1,1]$, is obtained from the recurrence relation: 

\begin{equation} \label{eq:def_cheb}
\begin{aligned}
T_0(x) &= 1, \quad T_1(x) = x, \\
T_{k+1}(x) &= 2xT_k(x)-T_{k-1}(x),
\end{aligned}
\end{equation}

Alternatively, the polynomials can be derived from the trigonometric definition, which gives exactly the same results:

\begin{equation} \label{eq:def_cheb_tri}
T_k (x) = T_k(\cos(\vartheta)) = \cos(k\vartheta).
\end{equation}

To use $T_n(x)$ as a representation for the position profile, a linear transformation from $t$ into the range $[-1,1]$ of $x$ is required \cite{Thompson2013}:

\begin{equation} \label{eq:rescale_tx}
\begin{aligned}
t&=\frac{1}{2}(t_B-t_A)x+\frac{1}{2}(t_B+t_A) =a x+ b,
\end{aligned}
\end{equation}

where scale factors $a$ and $b$ are defined for the purpose of the following paragraphs. In addition, the position $\theta \in [\theta_A, \theta_B]$ is also rescaled to the interval $\phi \in [-1,1]$, which makes it possible to obtain strict bounds on the design space in \eqref{boundsDS}. Thus, the rescaled motion profile description $\phi(x)$ of degree $n$ with optimizable coefficients $\mathbf{p} = [p_0, p_1, \ldots ,p_n]^T$ is obtained.

\begin{equation} \label{eq:position_function_cheb}
\phi(x)=\sum_{i=0}^{n} p_iT_i(x), \quad x\in [-1,1].
\end{equation}

The output of the motion simulations in the previous section deliver $n_s$ samples of inertia $\mathbf{J}= [J_1, \ldots  ,J_{n_s}]^T$, load torque $\boldsymbol{\uptau}_l= [\tau_{l,1}, \ldots  ,\tau_{l,n_s}]^T$ and corresponding angle query points $\boldsymbol{\uptheta} = [\theta_1, \ldots  ,\theta_{n_s}]^T$. Due to the position rescaling of the motion profile $\phi(x)$, the angle query points $\boldsymbol{\uptheta}$ have to be rescaled accordingly:

\begin{equation} \label{eq:rescale_prop}
\begin{aligned}
    \boldsymbol{\upphi} &= \frac{2}{(\theta_B -\theta_A)} \, \boldsymbol{\uptheta} - \frac{(\theta_B +\theta_A)}{(\theta_B -\theta_A)} = c \, \boldsymbol{\uptheta} + d.
    \end{aligned}
\end{equation}

Moreover, as the property description is now defined on the rescaled interval $\phi \in [-1,1]$, the following relationship holds with regard to the derivative properties such inertia variation $\frac{\mathrm{d}J(\phi)}{\mathrm{d}\phi}$:

\begin{equation} \label{eq:rescale_pder}
\frac{\mathrm{d}J(\phi)}{\mathrm{d}\phi} = \frac{1}{2}(\theta_B-\theta_A) \frac{\mathrm{d}J(\theta)}{\mathrm{d}\theta} =  e \,  \frac{\mathrm{d}J(\theta)}{\mathrm{d}\theta}.
\end{equation}

When using the rescaled position profile $\phi(x)$, it is important to rescale the torque equation \eqref{eq:torque_equation} as well. Otherwise, the resulting values of the torque profile $\tau(x)$ are distorted which results in different objective values (i.e. $\tau_{rms}$) and solutions. To preserve the motor torque's absolute values, the following rescaled torque equation is introduced:

\begin{equation} \label{eq:torque_equation_rescaled}
\tau_m(x) = \tau_l(\phi) + \frac{1}{2}\frac{\mathrm{d}J(\phi)}{\mathrm{d}\phi}\frac{1}{e}\left(\frac{\dot{\phi}}{a.c}\right)^2 + J(\theta)\frac{\ddot{\phi}}{a^2.c} + \mu_k\frac{\dot{\phi}}{a.c}.
\end{equation}

\begin{figure}[thpb]
  \centering	
  \includegraphics[width=\columnwidth]{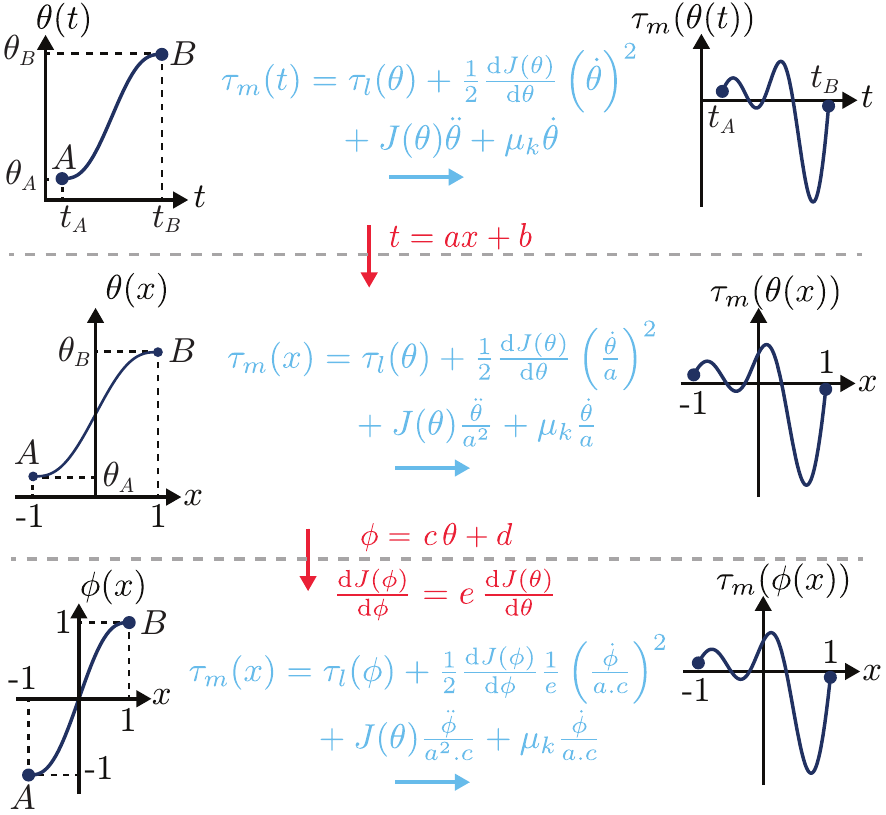}
  \caption{Original $\theta(t)$ and rescaled position profiles $\theta(x)$, $\phi(x)$ with their corresponding torque equations.}
  \label{fig:rescaling}
\end{figure}

An overview of the position and torque rescalings is presented in Fig. \ref{fig:rescaling}. The new system equation \eqref{eq:torque_equation_rescaled} ensures the system dynamics are equally scaled and the minima are not altered.

For what concerns the constraints, the rest-to-rest motion requires zero speed $\dot{\phi}$ and acceleration $\ddot{\phi}$ in the start and endpoint:

\begin{equation} 
\begin{array}{ccccc}
\phi(-1)=-1 & , & \dot{\phi}(-1)=0 & , & \ddot{\phi}(-1)=0,\\
\phi(1)=1 & , & \dot{\phi}(1)=0 & , & \ddot{\phi}(1)=0.
\end{array}{}
\label{eq:constraints}
\end{equation}

Referring to \eqref{eq:position_function_cheb}, and by incorporating the motion profile constraints \eqref{eq:constraints}, the lower degree coefficients $[p_0,... , p_5]^T$ can be written as a function of the remaining coefficients $[p_6,... , p_n]^T$, such that $n-5$ degrees of freedom (DOF) are kept available for the optimization algorithm \cite{Hsu2014}. Thus, the energy optimal motion profile problem is formulated as the following minimization problem with design variable vector $\mathbf{o}=[p_6,... , p_n]^T$:

\begin{equation}
\begin{aligned}
& \underset{\mathbf{o} \, \in \, \mathbb{R}^{n-5}}{\text{minimize}}
& & \tau_{rms} = \sqrt{\frac{1}{2}\int_{-1}^{1} {\tau_m(\phi(x,\mathbf{o}))}^2 \, \mathrm{d}x} .
\end{aligned}
\end{equation}


In some applications, an additional constraint of zero jerk in the begin and endpoint can be imposed to limit the vibrations:

\begin{equation} \label{eq:constraint_jerk}
    \dddot{\phi}(-1)=0 \quad;\quad \dddot{\phi}(1)=0.
\end{equation}

Because of these two extra equations, the DOF is reduced to $n-7$ and the design variable vector can be expressed as $\mathbf{o}=[p_8,... , p_n]^T$.

\subsection{Initialization \& Design Space}
In this paper, the resulting optimization problem is solved with both a fast \textit{gradient-based} solver, the BFGS (Broyden–Fletcher–Goldfarb–Shanno) quasi-Newton method \cite{Nocedal2006}, and a global \textit{heuristic} solver, the genetic algorithm \cite{Holland1992}.

For gradient-based optimization, a starting point needs to be defined. The use of the Chebyshev basis $T_i(x)$ in representation \eqref{eq:position_function_cheb} allows initializing the optimization parameter vector at zero since the coefficients in a convergent Chebyshev series development of the motion profile function $\phi(x)$ would converge to zero \cite{Majidian2017}. Here, we can safely assume some similar behavior for the coefficients $p_i$ in \eqref{eq:position_function_cheb}.

For what concerns the genetic algorithm, a similar approach is used for the initialization of the population. However, because a GA often samples a wide part of the design space \cite{Wenzhong2005}, it is beneficial to determine the exact bounds on the design vector $\mathbf{o}$. By doing so, the solver can cover a large part of the design space and reveal the global optimal solution. In the following paragraphs, thanks to the rescaled Chebyshev motion profile $\phi(x)$, strict bounds on the design vector $\mathbf{o}$ can be derived.

To define these bounds, we take a look at the projection of the position profile $\phi(x)$ onto the orthogonal Chebyshev polynomial basis $T_l(x)$. Given that $x =\cos(\theta)$, we introduce the inner product $F$:

\begin{equation}  \label{eq:integralF}
    \begin{aligned}
    F = \langle\phi(x),T_l(x)\rangle &= \int\limits_{-1}^{1} \frac{\phi(x)T_l(x)}{\sqrt{1-x^2}}\,\mathrm{d}x\\
    &= \int\limits_{0}^{2\pi} \phi(\cos\theta)T_l(\cos\theta)\,\mathrm{d}\theta.
    \end{aligned}
\end{equation}

Then, by taking into account the position function definition \eqref{eq:position_function_cheb}, we find the following result:

\begin{equation} 
    \begin{aligned}
    F &= \int\limits_{0}^{2\pi} \left(\sum\limits_{k=0}^n p_k T_k(\cos\theta)\right)T_l(\cos\theta)\,\mathrm{d}\theta \\
	&= \sum\limits_{k=0}^n p_k \int\limits_{0}^{2\pi} T_k(\cos\theta) T_l(\cos\theta) \, \mathrm{d}\theta.
    \end{aligned}
\end{equation}

Here, the integral $I= \int\limits_{0}^{2\pi} T_k(\cos\theta) T_l(\cos\theta) \, \mathrm{d}\theta$ can be further simplified by using the Chebyshev polynomial orthogonality properties, which are rederived here for the sake of readability. Because of Eq. \eqref{eq:def_cheb_tri} and by using the inverse Simpson rule of trigonometry, the integral $I$ can be written as:

\begin{equation}
\begin{aligned}
	I &= \int\limits_{0}^{2\pi} \cos(k\theta) \cos(\ell\theta) \, \mathrm{d}\theta \\
	&= \frac{1}{2} \int\limits_{0}^{2\pi} \cos\big((k+\ell)\theta\big)\,\mathrm{d}\theta
	+ \frac{1}{2} \int\limits_{0}^{2\pi} \cos\big((k-\ell)\theta\big)\,\mathrm{d}\theta.
\end{aligned}
\end{equation}

This integral can be split into three cases:
\begin{enumerate}
    \item \underline{$k =\ell = 0$} \\
    
    \begin{equation}\label{eq:case1}
        I = 
        2\pi,
    \end{equation}
    
    \item \underline{$k =\ell \neq 0$} \\
    
    \begin{equation}\label{eq:case2}
    \begin{aligned}
        I 
	= \pi,
	\end{aligned}
    \end{equation}
    
    \item \underline{$k \neq \ell$} \\
    
    \begin{equation}\label{eq:case3}
    \begin{aligned}
        I 
        = 0.
    \end{aligned}
    \end{equation}
\end{enumerate}

Thus, by taking into account \eqref{eq:case3}, only the term for which $k=l$ remains in the summation $F$:

\begin{equation}
    \begin{aligned}
    F = p_\ell \int\limits_{0}^{2\pi} \cos^2 (\ell\theta) \, \mathrm{d}\theta.
    \end{aligned}
\end{equation}

This can be split into two cases. For $\ell = 0$ and by making use of \eqref{eq:case1} and \eqref{eq:integralF} we find:
\begin{equation}\label{c0}
	p_0 = \dfrac{1}{2\pi} \int\limits_{0}^{2\pi} \phi(\cos\theta)\,\mathrm{d}\theta,
\end{equation}

and for $\ell > 0$, by making use of \eqref{eq:case2} and \eqref{eq:integralF}:

\begin{equation}\label{cl}
	p_\ell = \dfrac{1}{\pi} \int\limits_{0}^{2\pi} \phi(\cos\theta)\cos(\ell\theta)\,\mathrm{d}\theta.
\end{equation}

For $\theta \in [0,2\pi]$, $\cos\theta$ lies in interval $[-1,1]$. Because of the position rescalings of the motion profile $\phi(x)$, the image $\phi(\cos\theta)$ also lies in the interval $[-1,1]$. Thus, we find:

\begin{equation}
	|p_0| \leq \dfrac{1}{2\pi}\int\limits_{0}^{2\pi} |\phi(\cos\theta)|\,\mathrm{d}\theta \leq \dfrac{1}{2\pi}\int\limits_{0}^{2\pi} \,\mathrm{d}\theta = 1.
\end{equation}

and

\begin{equation}
|p_\ell| \leq \dfrac{1}{\pi}\int\limits_{0}^{2\pi} |\phi(\cos\theta)||\cos(\ell\theta)|\,\mathrm{d}\theta \leq \dfrac{1}{\pi}\int\limits_{0}^{2\pi}|\cos(\ell\theta)| \,\mathrm{d}\theta.
\end{equation}

To calculate this last integral, we use the periodicity of the function $\cos(\ell\theta)$. This function has a period of $2\pi / \ell$, so goes $\ell$ times up and down on the interval $[0,2\pi]$. So, after taking the absolute value of this function, we find $2\ell$ times the integral over the positive part of a period, for example, the interval $[-\pi/2\ell, \pi/2\ell]$:

\begin{equation}
\begin{aligned}
    \dfrac{1}{\pi}\int\limits_{0}^{2\pi}|\cos(\ell\theta)| \,\mathrm{d}\theta = \dfrac{2\ell}{\pi} \int\limits_{-\pi/2\ell}^{\pi/2\ell} \cos(\ell\theta)\,\mathrm{d}\theta = \dfrac{4}{\pi}.
\end{aligned}
\end{equation}

Thus, the following bounds for the coefficients $p_i$ are obtained:

\begin{equation}\label{boundsDS}
    |p_0| \leq 1 \hspace{0.5cm} \text{ and } \hspace{0.5cm} |p_\ell| \leq \dfrac{4}{\pi}, \hspace{0.3cm} \ell =1,\dots,n.
\end{equation}

These constraints on the design space simplify the subsequent optimization.

\section{Results}
\subsection{Motion Profile Optimization}
In order to assess the performance of the proposed method, a set of optimizations has been performed on the industrial pick-and-place unit depicted in Fig. \ref{fig:Experimental_Setup}. The mechanism is required to move between its start position $\theta_A$ of $0^\circ$ and end position $\theta_B$ of $173.6^\circ$ and has a motion time $\Delta t$ of $73.5ms$. As for the constraint, two different cases are considered, namely

\begin{itemize}
    \item \textit{Jerk Free (JF)}: Only the boundary constraints of \eqref{eq:constraints} are taken into account. The corresponding rescaled Chebyshev position profile $\phi(x)$ of degree $n$ is hereafter referred to as \textit{cheb"n"}. A 5th-degree polynomial, hereafter indicated as \textit{poly5}, is taken as the reference motion profile for comparison purposes. This is the smallest degree polynomial that satisfies the constraints. 
    
    \item \textit{Jerk Zero (J0)}: In addition to the constraint of a jerk-free optimization, a zero-jerk constraint is added in the start, and endpoint \eqref{eq:constraint_jerk} is added to the motion profile definition. The resulting $n$-th degree position profile $\phi(x)$ is referred to as \textit{cheb"n"J0}. The reference motion profile is in this case a 7th-degree polynomial, hereafter referred to as \textit{poly7J0}.
\end{itemize}

\begin{figure}[thpb]
  \centering	
  \includegraphics[width=\columnwidth]{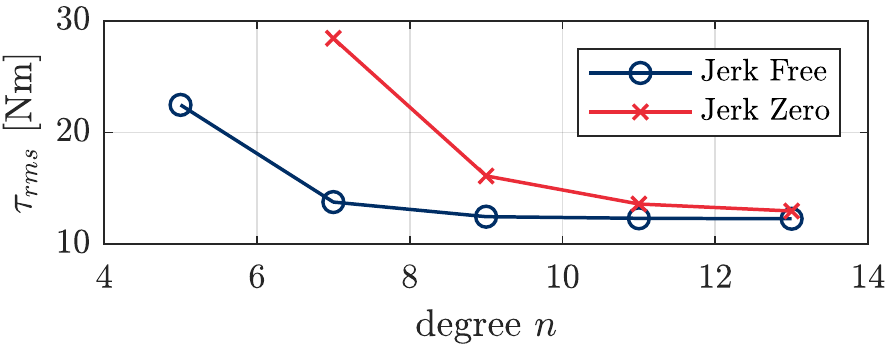}
  \caption{Results of the motion profile optimization for different degrees $n$.}
  \label{fig:Optimization_Results}
\end{figure}

For every case, the resulting optimization problem is solved in a MATLAB environment for degrees $n= 7, 9, 11,$ and $13$. The results are presented in Fig. \ref{fig:Optimization_Results} and Tables \ref{tab:results_jerkfree} \& \ref{tab:results_jerkzero} where for every motion profile, the corresponding RMS torque $\tau_{rms}$ and solve time $t_{sol}$ are displayed. Savings up to $54.4\%$ are obtained in under $0.77$ s. The results clearly converge towards a minimal value for increasing degree $n$. In general, the motion profiles which include the jerk constraint \eqref{eq:constraint_jerk}, have slightly bigger $\tau_{rms}$ values, which is to be expected due to the fact that this extra constraint limits the acceleration near the endpoints while it is desirable to have high accelerations here since the inertia is low.

In Table \ref{tab:results_jerkfree}, the $\tau_{rms}$ values of a conventional trapezoidal 1/3 motion profile are presented as well, which accelerates during 1/3rd of the time, moves at a constant speed during 1/3rd, and decelerates at the last 1/3rd \cite{Park1996}. What is interesting in this table is that the torque demand can already be significantly reduced by selecting an adequate default motion law. Notwithstanding that the greatest savings are realized after optimization.

It is worth noting that for the jerk-free motion profiles, the same solution was found for both the genetic algorithm and gradient-based solver. However, the calculation times with GA are considerably higher. When including the jerk constraint, the GA comes close but does not completely reveal the full optimization potential. Therefore, for what concerns the present study, gradient-based optimizations algorithms are preferable. Since the GA did not obtain a better solution for any motion profile in the bounded search space, we can expect that the results obtained with the gradient-based method are global optimal solutions.

Although only the forward motion is considered here, similar results can be obtained for the return motion by simply changing the position constraints.


\begin{table} 
\caption{Results of the motion profile optimization (Jerk Free).}
\label{tab:results_jerkfree}
\centering
\begin{tabular}{lcccc}
            & \multicolumn{2}{c}{\textbf{Gradient-Based }}                                       & \multicolumn{2}{c}{\textbf{Genetic Algorithm}}                                    \\
\textbf{JF} & \textbf{$\tau_{rms}\,[Nm]$}                                & \textbf{$t_{sol}\, [s]$} & \textbf{\textbf{$\tau_{rms}\,[Nm]$}}                      & \textbf{$t_{sol}\,[s]$}  \\ 
\hline\hline
poly5 (ref.)       & \begin{tabular}[c]{@{}c@{}}22.48\end{tabular}        & -                        & \begin{tabular}[c]{@{}c@{}}22.48\end{tabular}       & -                        \\ 
\cmidrule(lr){1-5}
trap       & \begin{tabular}[c]{@{}c@{}}17.16 \\-23.7\%\end{tabular} & -                     & \begin{tabular}[c]{@{}c@{}}17.16\\-23.7\%\end{tabular} & -                     \\ 
\cmidrule(lr){1-5}
cheb7       & \begin{tabular}[c]{@{}c@{}}13.78 \\-38.7\%\end{tabular} & 0.21                     & \begin{tabular}[c]{@{}c@{}}13.78\\-38.7\%\end{tabular} & 3.28                     \\ 
\cmidrule(lr){1-5}
cheb9       & \begin{tabular}[c]{@{}c@{}}12.47 \\-44.5\%\end{tabular} & 0.32                     & \begin{tabular}[c]{@{}c@{}}12.47\\-44.5\%\end{tabular} & 40.33                    \\ 
\cmidrule(lr){1-5}
cheb11      & \begin{tabular}[c]{@{}c@{}}12.33 \\-45.2\%\end{tabular} & 0.51                     & \begin{tabular}[c]{@{}c@{}}12.33\\-45.2\%\end{tabular} & 67.05                    \\ 
\cmidrule(lr){1-5}
cheb13      & \begin{tabular}[c]{@{}c@{}}12.29 \\-45.4\%\end{tabular} & 1.06                     & \begin{tabular}[c]{@{}c@{}}12.29\\-45.4\%\end{tabular} & 142.34                   \\
\cmidrule(lr){1-5}
\end{tabular}
\end{table}

\begin{table} 
\caption{Results of the motion profile optimization (Jerk 0).}
\label{tab:results_jerkzero}
\centering
\begin{tabular}{lcccc}
            & \multicolumn{2}{c}{\textbf{Gradient-Based }}                                       & \multicolumn{2}{c}{\textbf{Genetic Algorithm}}                                    \\
\textbf{J0} & \textbf{$\tau_{rms}\,[Nm]$}                                & \textbf{$t_{sol}\, [s]$} & \textbf{\textbf{$\tau_{rms}\,[Nm]$}}                      & \textbf{$t_{sol}\,[s]$}  \\ 
\hline\hline
poly7J0 (ref.)     & \begin{tabular}[c]{@{}c@{}}28.44\end{tabular}        & -                        & \begin{tabular}[c]{@{}c@{}}28.44\end{tabular}       & -                        \\ 
\cmidrule(r){1-5}
cheb9J0     & \begin{tabular}[c]{@{}c@{}}16.12 \\-43.3\%\end{tabular} & 0.27                     & \begin{tabular}[c]{@{}c@{}}16.12\\-43.3\%\end{tabular} & 6.15                     \\ 
\cmidrule(r){1-5}
cheb11J0    & \begin{tabular}[c]{@{}c@{}}13.61 \\-52.2\%\end{tabular} & 0.38                     & \begin{tabular}[c]{@{}c@{}}14.11\\-50.4\%\end{tabular} & 175.23                   \\ 
\cmidrule(lr){1-5}
cheb13J0      & \begin{tabular}[c]{@{}c@{}}12.98\\-54.4\%\end{tabular}  & 0.77                     & \begin{tabular}[c]{@{}c@{}}13.15\\-53.8\%\end{tabular} & 195.02                   \\
\cmidrule(lr){1-5}
\end{tabular}
\end{table}

\subsection{Measurements}
\begin{figure}[thpb]
  \centering	
  \includegraphics[width=\columnwidth]{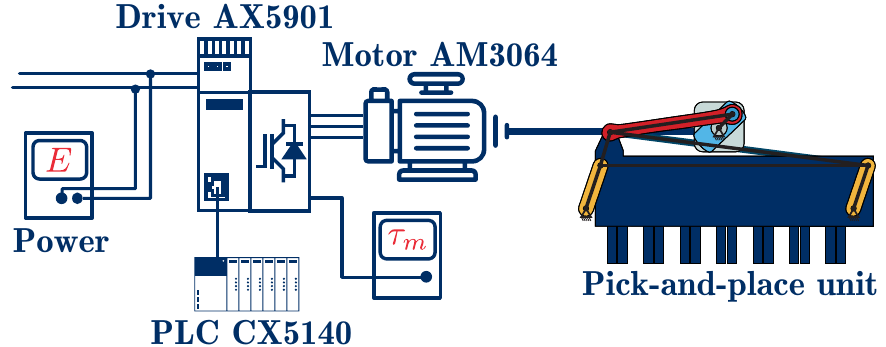}
  \caption{Schematic overview of the experimental setup.}
  \label{fig:Schematic_Experimental_Setup}
\end{figure}

The theoretical results are validated against experimental measurements on the pick-and-place unit (Fig. \ref{fig:Experimental_Setup}). The setup comprises a Beckhoff CX5140 PLC, a Beckhoff AX5901 motor drive, and a Beckhoff AM3064 PMSM, which is connected to the shaft of the mechanism. In order to measure the input electrical energy, a Tektron PA4000 power analyzer is used to analyze the power supply (Fig. \ref{fig:Schematic_Experimental_Setup}).

The theoretical savings potential of the motion profile optimization is only fulfilled when the motor is capable of following the optimized position setpoint. Therefore, a performant motion controller needs to be designed in order to keep the tracking error as low as possible. Here, similar to \cite{VanOosterwyck2019}, a cascade controller with torque and speed feedforward is employed as it has proven to be successful for high dynamic systems. The look-up table for the feedforward torque is determined using the torque equation \eqref{eq:torque_equation_rescaled}.

\begin{figure}[thpb]
  \centering	
  \includegraphics[width=\columnwidth]{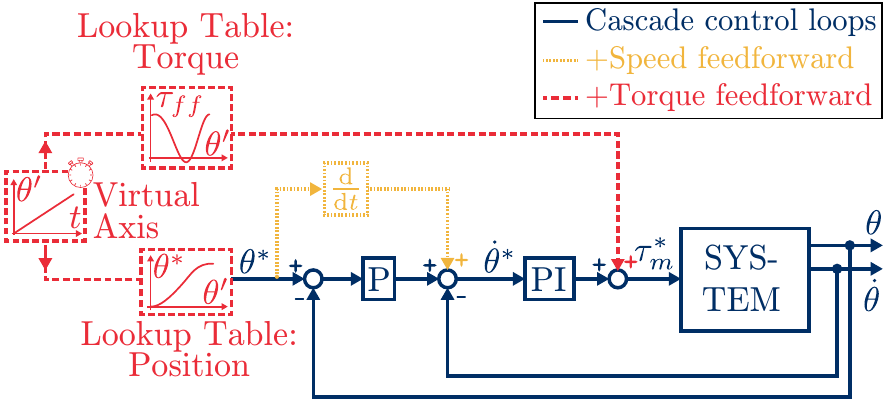}
  \caption{Schematic overview of the cascade motion controller with feedforward \cite{VanOosterwyck2019}.}
  \label{fig:Motion_Controllers}
\end{figure}

In Tables \ref{tab:measurement_jerkfree} and \ref{tab:measurement_jerkzero}, the results of both the measured RMS torque $\tau_{rms}$ and measured input electrical energy $E$ for different motion profiles are presented. As expected from the simulations, the lowest absolute energy consumption is obtained when using jerk-free motion profiles. When the jerk constraint is active, a decrease of 62.9\% in energy consumption can be achieved by optimizing the motion profile, while a relative saving of 52.5\% is possible if no extra constraint on the jerk is imposed.

The measured $\tau_{rms;meas}$ and calculated RMS motor torque $\tau_{rms}$ show a very high similarity, which confirms that the present system model is valid.

\begin{table}
\caption{Experimental results with energy measurement (Jerk Free).}
\label{tab:measurement_jerkfree}
\centering
\begin{tabular}{cccc}
\textbf{JF} & \textbf{$\tau_{rms}\,[Nm]$}                               & \textbf{$\tau_{rms;meas}\,[Nm]$}                          & \textbf{$E_{meas} \, [Wh] $}                                   \\ 
\hline\hline
poly5       & 22.48                                                  & 19.59                                                  & 312.2                                                   \\ 
\cmidrule(r){1-4}
trap       & \begin{tabular}[c]{@{}c@{}}17.16\\-23.7\%\end{tabular} & \begin{tabular}[c]{@{}c@{}}15.88\\-18.98\%\end{tabular} & \begin{tabular}[c]{@{}c@{}}215.1\\-31.1\%\end{tabular}  \\ 
\cmidrule(r){1-4}
cheb7       & \begin{tabular}[c]{@{}c@{}}13.78\\-38.7\%\end{tabular} & \begin{tabular}[c]{@{}c@{}}13.40\\-31.6\%\end{tabular} & \begin{tabular}[c]{@{}c@{}}181.7\\-41.8\%\end{tabular}  \\ 
\cmidrule(lr){1-4}
cheb9       & \begin{tabular}[c]{@{}c@{}}12.47\\-44.5\%\end{tabular} & \begin{tabular}[c]{@{}c@{}}12.07\\-38.4\%\end{tabular} & \begin{tabular}[c]{@{}c@{}}152.3\\-51.2\%\end{tabular}  \\ 
\cmidrule(lr){1-4}
cheb11      & \begin{tabular}[c]{@{}c@{}}12.33\\-45.2\%\end{tabular} & \begin{tabular}[c]{@{}c@{}}11.93\\-39.1\%\end{tabular} & \begin{tabular}[c]{@{}c@{}}150.1\\-51.9\%\end{tabular}  \\ 
\cmidrule(lr){1-4}
cheb13      & \begin{tabular}[c]{@{}c@{}}12.29\\-45.4\%\end{tabular} & \begin{tabular}[c]{@{}c@{}}11.83\\-39.6\%\end{tabular} & \begin{tabular}[c]{@{}c@{}}148.2\\-52.5\%\end{tabular}  \\
\cmidrule(lr){1-4}
\end{tabular}
\end{table}

\begin{table}
\caption{Experimental results with energy measurement (Jerk Zero).}
\label{tab:measurement_jerkzero}
\centering
\begin{tabular}{cccc}
\textbf{J0} & \textbf{$\tau_{rms}\,[Nm]$}                               & \textbf{$\tau_{rms;meas}\,[Nm]$}                          & \textbf{$E_{meas} \, [Wh] $}                                   \\ 
\hline\hline
poly7J0     & 28.44                                                  & 25.30                                                  & 458.5                                                   \\ 
\cmidrule(lr){1-4}
cheb9J0     & \begin{tabular}[c]{@{}c@{}}16.12\\-43.3\%\end{tabular} & \begin{tabular}[c]{@{}c@{}}15.81\\-37.5\%\end{tabular} & \begin{tabular}[c]{@{}c@{}}222.9\\-51.4\%\end{tabular}  \\ 
\cmidrule(lr){1-4}
cheb11J0    & \begin{tabular}[c]{@{}c@{}}13.61\\-52.2\%\end{tabular} & \begin{tabular}[c]{@{}c@{}}13.08\\-48.3\%\end{tabular} & \begin{tabular}[c]{@{}c@{}}170.3\\-62.9\%\end{tabular}  \\ 
\cmidrule(lr){1-4}
cheb13J0    & \begin{tabular}[c]{@{}c@{}}12.98\\-54.4\%\end{tabular} & \begin{tabular}[c]{@{}c@{}}12.72\\-49.7\%\end{tabular} & \begin{tabular}[c]{@{}c@{}}170.8\\-62.7\%\end{tabular}  \\
\cmidrule(lr){1-4}
\end{tabular}
\end{table}
\section{Conclusion}
This study proposes a novel approach for motion profile optimization of PTP motions with Chebyshev polynomials. At first, system properties have been extracted from both CAD motion simulations and measurements to obtain an accurate virtual twin of the system. A Chebyshev motion profile with scaling laws is presented. Especially novel in this paper is the derivation of the boundary conditions of this profile which enables to define bounds for the design variables. The latter allows to use an optimizer that is designed to obtain globally optimal solutions, i.e. Genetic Algorithm. In addition, the solutions are validated with fast gradient-based optimization algorithms. Finally, experimental optimization results have been considered to verify the feasibility of the proposed solutions.

The numerical results, achieved on an exemplary model, clearly show that large $\tau_{rms}$ savings of up to 53.8\% can be achieved. In addition, it is shown that by employing Chebyshev polynomials for the motion profile, a fast gradient-based optimization can be effectively employed with solve times under $0.8$s. At last, the validation measurements show that similar savings are obtained on the real machine with a maximum energy reduction of $62.9 \%$.

Due to the straightforward implementation of both the optimization itself and integration of the resulting motion profiles in the motor drive, the proposed method can be easily adopted in any existing configuration where the CAD is data available. Therefore, the proposed method is expected to have a beneficial impact on the energy usage of the envisaged PTP applications.

\section*{Acknowledgements}
Research funded by a PhD grant of the Research Foundation Flanders (FWO) [1S88120N].

\bibliography{references}

\end{document}